\documentclass[11pt,a4paper]{article}
\pdfoutput=1
\usepackage{jheppub}
\usepackage{microtype}
\usepackage{dcolumn}
\usepackage{booktabs}

\usepackage{siunitx}
\usepackage{xspace}

\newcommand{\NEW}{NEXT-White}
\newcommand{\NEXT}{NEXT-100}

\newcommand{\Fig}{Figure}

\newcommand{\eg}{{\it e.g.}}

\newcommand{\bb}{\ensuremath{\beta\beta}}

\newcommand{\bbonu}{\ensuremath{\beta\beta0\nu}}

\newcommand{\tz}{\ensuremath{t_0}}
\newcommand{\stwo}{\ensuremath{S_2}}
\newcommand{\sone}{\ensuremath{S_1}}

\newcommand{\mbb}{\ensuremath{m_{\beta\beta}}}

\newcommand{\ckky}{\ensuremath{\rm counts~keV^{-1}~kg^{-1}~yr^{-1})}}

\newcommand{\Qbb}{\ensuremath{Q_{\beta\beta}}}

\newcommand{\Tonu}{\ensuremath{T_{1/2}^{0\nu}}}

\newcommand{\CS}{\ensuremath{^{137}}Cs}

\newcommand{\RAD}{\ensuremath{^{222}}Rn}

\newcommand{\XE}{\ensuremath{{}^{136}\rm Xe}}

\newcommand{\TL}{\ensuremath{{}^{208}\rm{Tl}}}

\newcommand{\THO}{\ensuremath{{}^{232}{\rm Th}}}

\newcommand{\BI}{\ensuremath{{}^{214}}Bi}
\newcommand{\Bapp}{Ba\ensuremath{^{++}}}

\newcommand{\Tl}[1]{\ensuremath{^{#1}\mathrm{Tl}}\xspace}

\newcommand{\Xe}[1]{\ensuremath{^{#1}\mathrm{Xe}}\xspace}

\DeclareSIUnit\c{\mbox{$c$}}
\DeclareSIUnit\magn{\mbox{$\times$}}
\DeclareSIUnit\min{min}
\DeclareSIUnit\week{week}
\DeclareSIUnit\year{yr}
\DeclareSIUnit\years{years}
\DeclareSIUnit\yr{yr}
\DeclareSIUnit\standard{std}
\DeclareSIUnit\str{sr}
\DeclareSIUnit\ppm{ppm}
\DeclareSIUnit\ppb{ppb}
\DeclareSIUnit\ppt{ppt}
\DeclareSIUnit\pe{PE}
\DeclareSIUnit\spe{SPE}
\DeclareSIUnit\ev{events}
\DeclareSIUnit\ct{counts}
\DeclareSIUnit\neutron{\mbox{$n$}}
\DeclareSIUnit\smp{samples}
\DeclareSIUnit\Sample{S}
\DeclareSIUnit\ch{ch}
\DeclareSIUnit\hit{hit}
\DeclareSIUnit\hits{hits}
\DeclareSIUnit\bin{(\mbox{5-PE}~bin)}
\DeclareSIUnit\sgm{\mbox{$\sigma$}}
\DeclareSIUnit\rms{RMS}
\DeclareSIUnit\keVr{\mbox{keV$_{\rm nr}$}}
\DeclareSIUnit\keVee{\mbox{keV$_{e{\rm e}}$}}
\DeclareSIUnit\ph{photon}
\DeclareSIUnit\pes{pes}
\DeclareSIUnit\el{electrons}
\DeclareSIUnit\pm{PMT}
\DeclareSIUnit\inch{"}
\DeclareSIUnit\bit{bit}
\DeclareSIUnit\sample{samples}
\DeclareSIUnit\barn{barn}
\DeclareSIUnit\bara{bar}
\DeclareSIUnit\barg{barg}
\DeclareSIUnit\mlardepth{\mbox(meter~of~\LAr~depth)}
\DeclareSIUnit\Curie{Ci}
\DeclareSIUnit\psi{psi}
\DeclareSIUnit\parsec{pc}
\DeclareSIUnit\liveday{\mbox{live-days}}
\DeclareSIUnit\days{\mbox{days}}
\DeclareSIUnit\day{\mbox{day}}
\DeclareSIUnit\miles{\mbox{miles}}
\DeclareSIUnit\degreeC{\mbox{$^{\circ}$C}}
\DeclareSIUnit\electron{\mbox{$e^-$}}
\DeclareSIUnit\Euro{\mbox{\euro}}
\DeclareSIUnit\cph{cph}
\DeclareSIUnit\neq{neq}
\DeclareSIUnit\unit{unit}
\DeclareSIUnit\byte{Byte}
\DeclareSIUnit\Bq{\becquerel}

\newcommand{\XeWaveLength}{\SI{172}{\nano\meter}}

\newcommand{\Next}{\mbox{NEXT-100}}

\newcommand{\Ntk}{\mbox{NEXT-2.0}}

\newcommand{\HPXeEL}{HPXe-EL}

\newcommand{\TPB}{Tetraphenyl Butadiene}
\newcommand{\PDOT}{Poly-Ethylenedioxythiophene}

\newcommand{\XenonQbb}{\SI{2458}{\keV}}

\newcommand{\Z}{\ensuremath{z}}

\newcommand{\NewPressureVesselMaterial}{316Ti}

\newcommand{\TypicalReducedField}{\SI{2}{\kV\per\cm\per\bar}}

\newcommand{\NewNumberOfSiPM}{\num{1792}}

\newcommand{\NewNumberOfBoards}{\num{28}}
\newcommand{\NewNumberOfSiPMPerBoard}{8 x 8}
\newcommand{\NewSiPMSeries}{SensL~C}
\newcommand{\NewSiPMModel}{MicroFC-10035-SMT-GP}
\newcommand{\NewSiPMSize}{\SI{1}{\mm\square}}
\newcommand{\NewSipmPitch}{\SI{10}{\mm}}

\newcommand{\TrackingPlaneToEL}{\SI{8}{\mm}}

\newcommand{\NewTypePMT}{Hamamatsu R11410-10}

\newcommand{\XeEnrichment}{\SI{90}{\percent}}

\newcommand{\NextTpcDiameter}{\SI{1050}{\mm}}
\newcommand{\NextTpcLength}{\SI{1300}{\mm}}
\newcommand{\NextFiducialVolume}{\SI{1.27}{\cubic\meter}}

\newcommand{\NextFiducialMass}{\SI{97}{\kg}}

\newcommand{\NextPressure}{\SI{15}{\bar}}

\newcommand{\NextNumberOfSiPM}{\num{5600}}
\newcommand{\NextNumberOfPMT}{\num{60}}

\begin{document}
\title{Status and prospects of the NEXT experiment for neutrinoless double beta decay searches}

\author{ J.J. Gomez-Cadenas }

\affiliation{Donostia International Physics Center (DIPC) and IKERBASQUE,\\
Paseo Manuel de Lardizabal, 4
20018 Donostia-San Sebasti\'an, Gipuzkoa, Spain}

\abstract{
NEXT (Neutrino Experiment with a Xenon TPC) is an experimental program whose goals are to discover neutrinoless double beta decay using \XE\ in high pressure xenon TPCs with electroluminescent readout. In this paper, results from the \NEW\ detector, which is currently taking data at Laboratorio Subterr\'aneo de Canfranc (LSC) will be reported. The prospects for the \Next\ apparatus, scheduled to start operations in 2020, as well as the plans to extend the technology to large and ultra-low background detectors needed to fully explore the inverse hierarchy of neutrino masses, will also be briefly discussed. }

\maketitle

\clearpage
\section{Introduction}

The NEXT program is developing the technology of high-pressure xenon gas time projection chambers (TPCs) with electroluminescent amplification (\HPXeEL) for neutrinoless double beta decay searches (\bbonu)~\citep{Nygren:2009zz, Gomez-Cadenas:2013lta, Martin-Albo:2015rhw}. The first phase of the program included the construction, commissioning and operation of two prototypes, called NEXT-DEMO and NEXT-DBDM (with masses of around 1 kg), which demonstrated the robustness of the technology, its excellent energy resolution and its unique topological signal \citep{Alvarez:2012xda, Alvarez:2013gxa, Alvarez:2012hh, Ferrario:2015kta}. 
The \NEW \footnote{Named after Prof.~James White, our late mentor and friend.} demonstrator, deploying 5~kg of xenon, implements the second phase of the program. \Next\  constitutes the third phase of the program. It is a radiopure detector deploying 100~kg of xenon at 15~bar and scaling up \NEW\ by slightly more than 2:1 in linear dimensions. The fourth phase of the program will explore two possible approaches to build ton-scale, ultra-low background detectors, called respectively, ``high definition'' (NEXT-HD) and ``Barium iOn Light Detection'' (NEXT-BOLD).  NEXT-HD will optimize the energy resolution and topological signature already proven by the technology, while reducing drastically the radioactive budget; NEXT-BOLD would be a detector implementing a delayed coincidence between the observation of the electron ionization signal and the observation of the single barium ion produced in a double beta decay. This paper reports the results obtained so far with \NEW\ (section \ref{sec:NEW}), describes the prospects of  \Next, which is scheduled to start operations in 2020 (section \ref{sec.next100}), and briefly sketches the prospects of the envisioned next-generation apparatus, NEXT-HD and  NEXT-BOLD (section \ref{sec.ntk}). Conclusions and outlook will be presented in section \ref{sec.conclu}.

\section{The \NEW\  demonstrator}
\label{sec:NEW}

\subsection{Principle of operation of \HPXeEL\ TPCs}
\label{sec.hpxe}

\begin{figure}[bhtp!]
\centering
\includegraphics[width=0.55\textwidth]{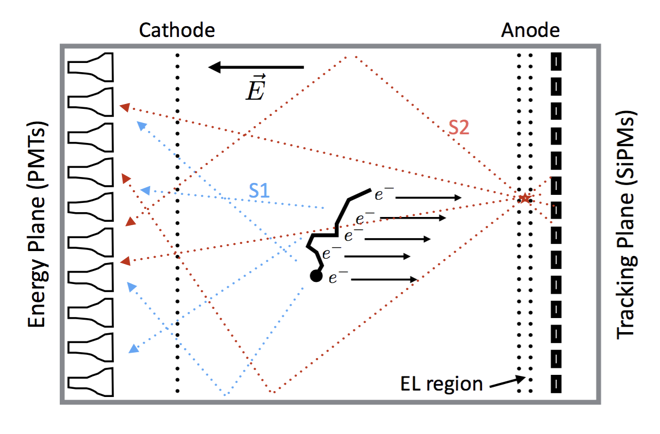}
\caption{\small Principle of operation of an \HPXeEL.} 
\label{fig.po}
\end{figure}

\Fig\ \ref{fig.po} illustrates the principle of operation of a high pressure xenon gas TPC with electroluminescent amplification of the ionization signal (\HPXeEL\ TPC). Charged tracks propagating in the TPC
lose energy by ionizing and exciting the gas atoms. The latter will de-excite by emitting vacuum ultraviolet (VUV) photons of $\sim$\XeWaveLength\ wavelength. This prompt scintillation is the so-called primary signal (\sone). 
The electrons produced by ionization along the track are drifted, under the influence of an electric field, towards the TPC anode, where they are amplified using electroluminescence (EL). This EL is produced using a more intense electric field (of order \TypicalReducedField) between the TPC gate, which is held at high voltage, and the grounded anode. This field strength accelerates the ionization electrons sufficiently to produce \stwo\ scintillation light.  In \NEW\ (and in \Next), the light is detected by two independent sensor planes located behind the anode and the cathode. The energy of the event is measured by integrating the amplified EL signal (\stwo) with a plane of photomultipliers (PMTs). This {\em energy plane} also records the \sone\ signal which provides the start-of-event (\tz).  
EL light is also detected a few mm away from production at the anode plane by a dense array of silicon photomultipliers (SiPMs), known as the  \emph{tracking plane}. This measurement allows for topological reconstruction since it provides position information transverse to the drift direction. The longitudinal (or \Z) position of the event is obtained using the drift time which is defined as the time difference between the \sone\ and \stwo\ signals.

\subsection{Demonstration of the NEXT technology with \NEW}
\label{sec.new}

\begin{figure}[bhtp!]
\centering
\includegraphics[width=0.49\textwidth]{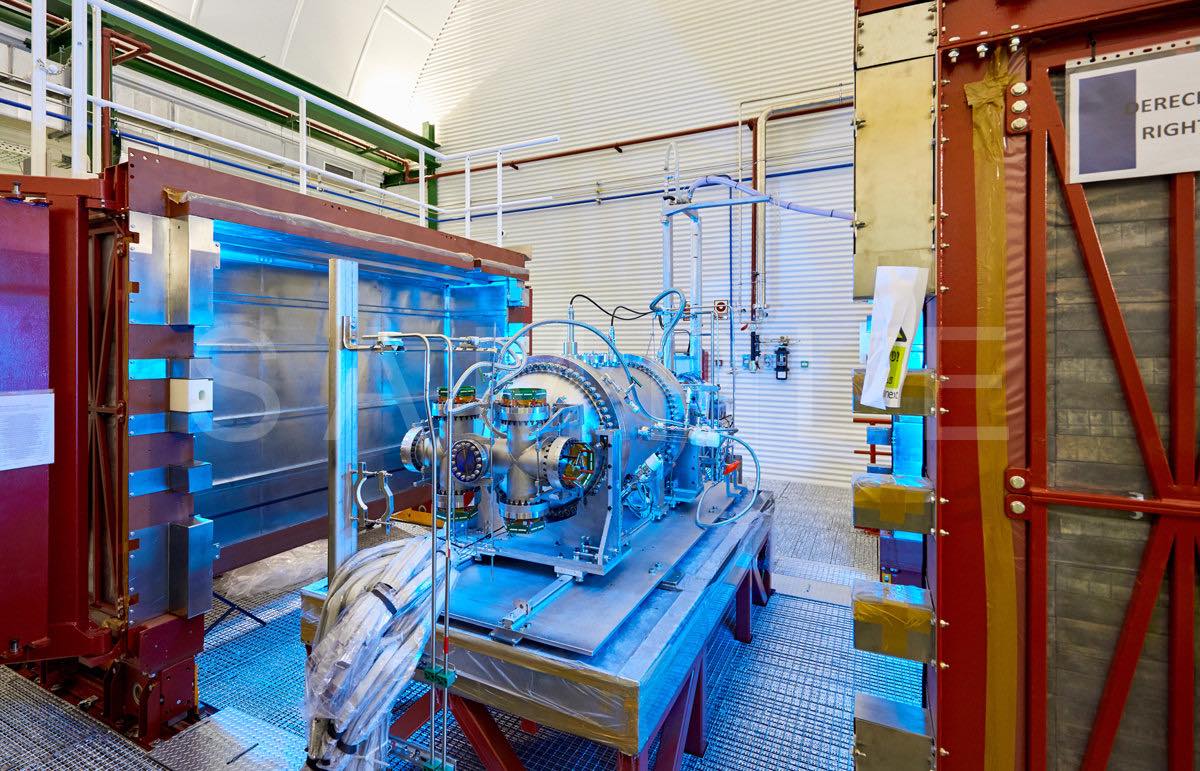}
\includegraphics[width=0.49\textwidth]{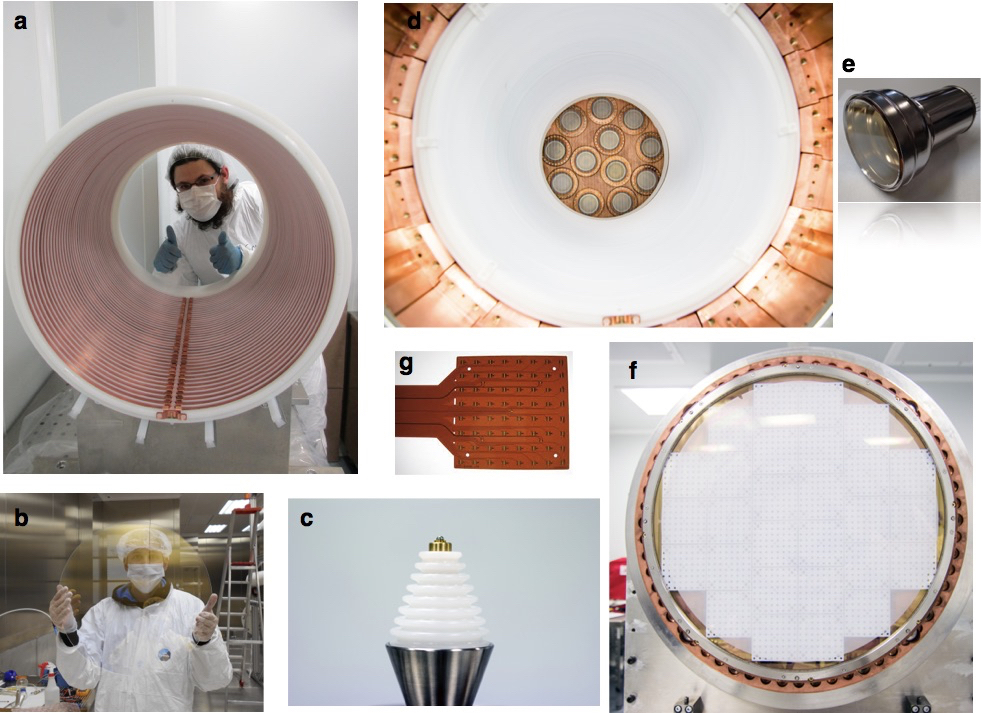}
\caption{\small Left: the \NEW\ detector at the LSC. Right: a selection of the main subsystems of \NEW: a) the field cage; b) the anode plate; c) high voltage feedthrough; d) energy plane; e) PMTs used in the energy plane; f) tracking plane; g) kapton boards composing the tracking plane.} 
\label{fig.new2}
\end{figure} 

The \NEW\ apparatus \citep{Monrabal:2018xlr}, shown in figure.~\ref{fig.new2}, operates inside a pressure vessel fabricated with radiopure stainless steel alloy, \NewPressureVesselMaterial. The pressure vessel is surrounded by a lead shield. Since a long electron lifetime is mandatory, the xenon circulates in a sophisticated gas system. The whole setup is located in Hall-A of Laboratorio Subterr\'aneo de Canfranc (LSC).  The right panel of figure ~\ref{fig.new2} shows a selection of the main subsystems of \NEW. The main subdetectors are the time projection chamber (TPC), the electroluminescent (EL) region, the energy plane and the tracking plane.

The field cage body is a High Density Polyethylene (HDPE) cylinder. Radiopure copper rings are inserted in the inner part of the field cage body (figure ~\ref{fig.new2}-a). The drift field is created by applying a voltage difference between the cathode and the gate, through high voltage feedthroughs (Fig.~\ref{fig.new2}-c). The field transports ionization electrons to the anode where they are amplified.
 
The amplification or electroluminescent region is the most delicate part of the detector, given the requirements for a high and yet very uniform electric field. The anode is defined by a fused silica plate coated with \PDOT\ (PEDOT) (figure~\ref{fig.new2}-b). A  layer of \TPB\ (TPB), commonly used in noble gases detectors to shift vaccuum ultraviolet (VUV) light to the visible spectrum, is vacuum-deposited on top of the PEDOT.

The measurement of the event energy as well as the detection of the primary scintillation signal that determines the \tz\ of the event is performed by the energy plane (EP), shown in figure~\ref{fig.new2}-d. The \NewTypePMT\ PMTs (figure~\ref{fig.new2}-e) were chosen for their low radioactivity \citep{Cebrian:2017jzb} and good performance. Since they cannot withstand high pressure they are protected from the main gas volume by a radiopure copper plate, which also acts as a shielding against external radiation. The PMTs are coupled to the xenon gas volume through sapphire windows. The windows are coated with PEDOT. A thin layer of TPB is vacuum-deposited on top of the PEDOT.

The tracking function is performed by a plane holding a sparse matrix of SiPMs. The sensors have a size of \NewSiPMSize\ and are placed at a pitch of \NewSipmPitch. The tracking plane is placed behind the anode with a total distance to the center of the EL region of \TrackingPlaneToEL. The sensors are \NewSiPMSeries\ series model \NewSiPMModel. The SiPMs are distributed in \NewNumberOfBoards\ boards (DICE boards) with \NewNumberOfSiPMPerBoard\ pixels each for a total of \NewNumberOfSiPM\ sensors (figure~\ref{fig.new2}-g). The DICE boards are mounted on a copper plate intended to shield against external radiation. The material used for the DICE boards is a low-radioactivity kapton printed circuit. The \NEW\ tracking plane (figure~\ref{fig.new2}-f) is currently the largest system deploying SiPMs as light pixels in the world.

\subsubsection*{A selection of results from the \NEW\ detector}
\label{sec.operationRunII}

\NEW\ started operations in late fall, 2016. After a short engineering run (Run I) in November-December 2016, the detector has been operating continuously.  The detector operated with \XE-depleted xenon for 7 months in 2017 at a pressure of $\sim$\SI{7}{~\bar} (Run II), and for 9 months in 2018 at a pressure of \SI{10}{~\bar} (Run IV). Run V,  with \XE-enriched xenon, is currently active. Operation in Run II established a procedure to calibrate the detector with krypton decays \citep{Martinez-Lema:2018ibw}, and provided initial measurements of energy resolution \citep{Renner:2018ttw}, electron drift parameters such as drift velocity, and transverse and longitudinal diffusion \citep{Simon:2018vep} and a measurement of the impact of \RAD\ in the radioactive budget, that was found to be small \citep{Novella:2018ewv}.


\begin{figure}[h!]
\centering

\includegraphics[width=0.58\textwidth]{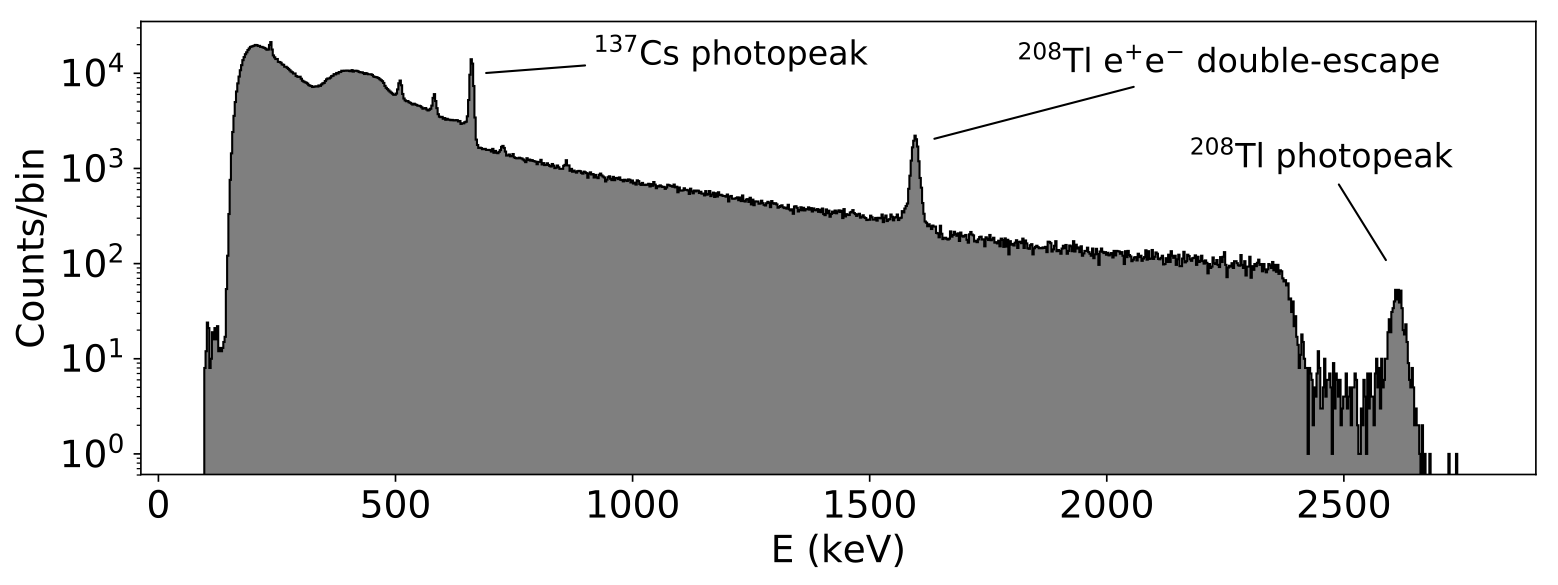}
\includegraphics[width=0.32\textwidth]{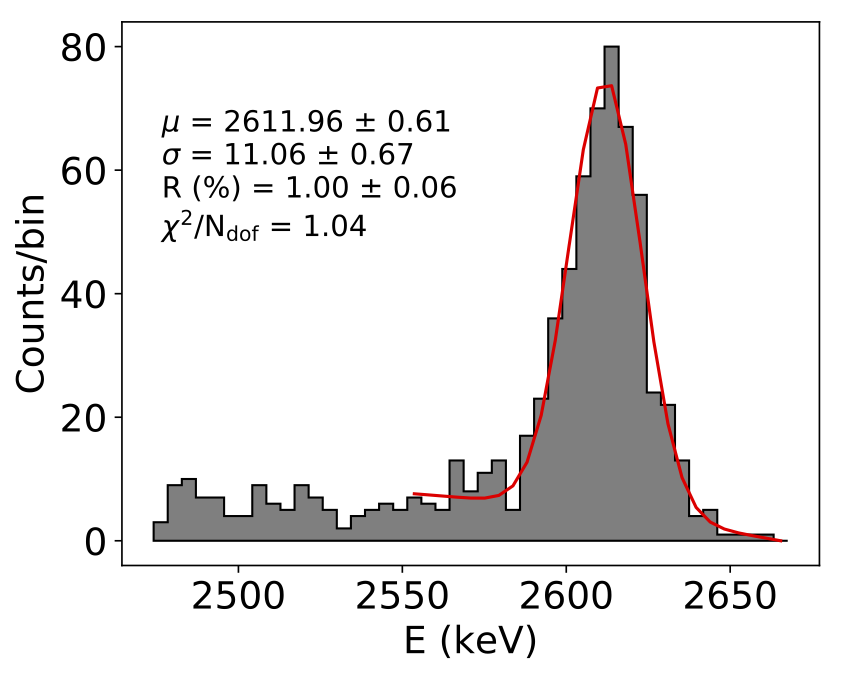}
\caption{Energy resolution measured by \NEW. Left panel: The full calibration spectrum obtained with \CS\ and \THO\ sources; right panel: energy resolution at the \THO\ photopeak.}
\label{fig.csth}
\end{figure}

\begin{figure}[h!]
\centering
\includegraphics[width=0.50\textwidth]{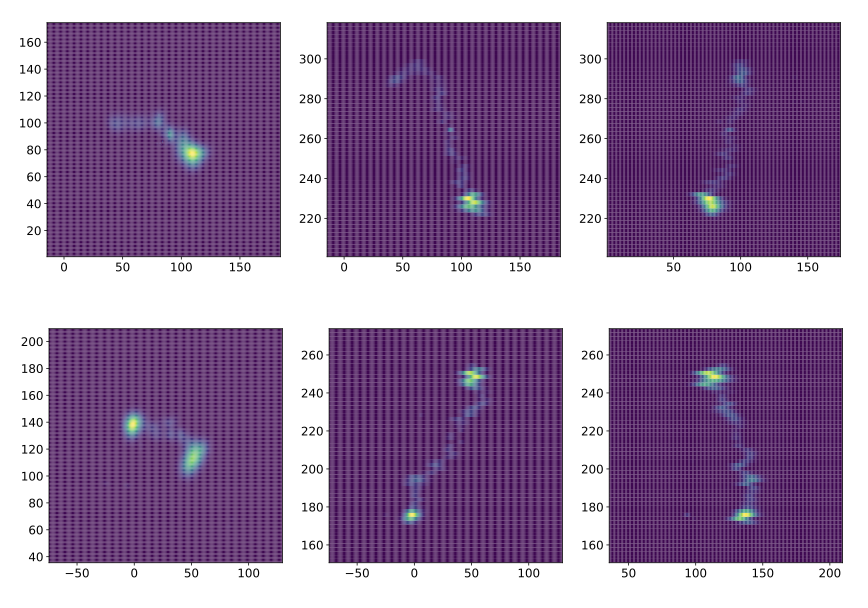}
\includegraphics[width=0.40\textwidth]{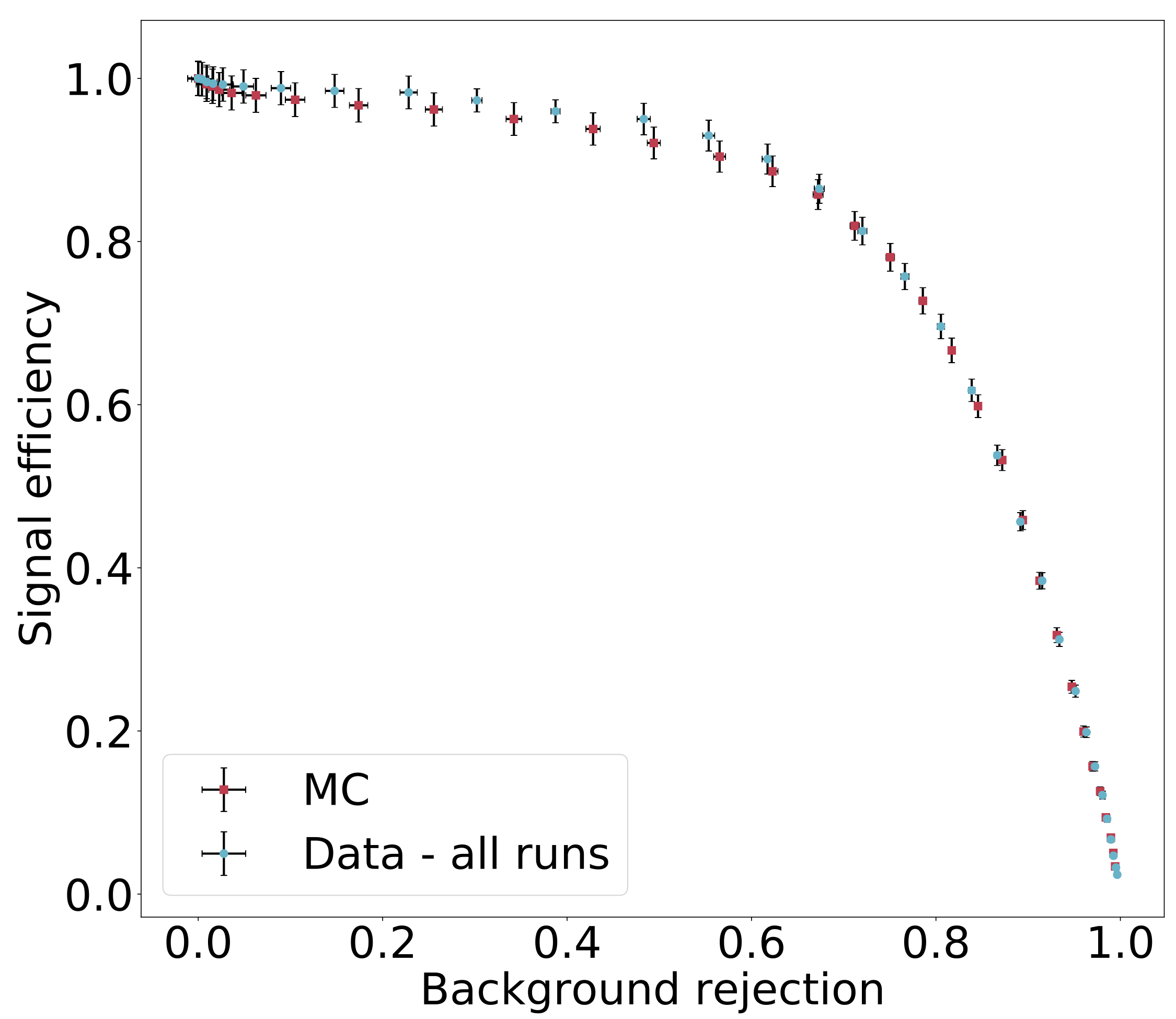}
\caption{Topological discrimination of signal and background measured by \NEW. Left panel: examples of double and single electrons with energies in the region of the \THO\ double escape peak from data; right: topological discrimination of signal and background at the energies of the \THO\ double escape peak, showing good agreement between data and Monte Carlo.}
\label{fig.topo}
\end{figure}


\begin{figure}[h!]
\centering
\includegraphics[width=0.43\textwidth]{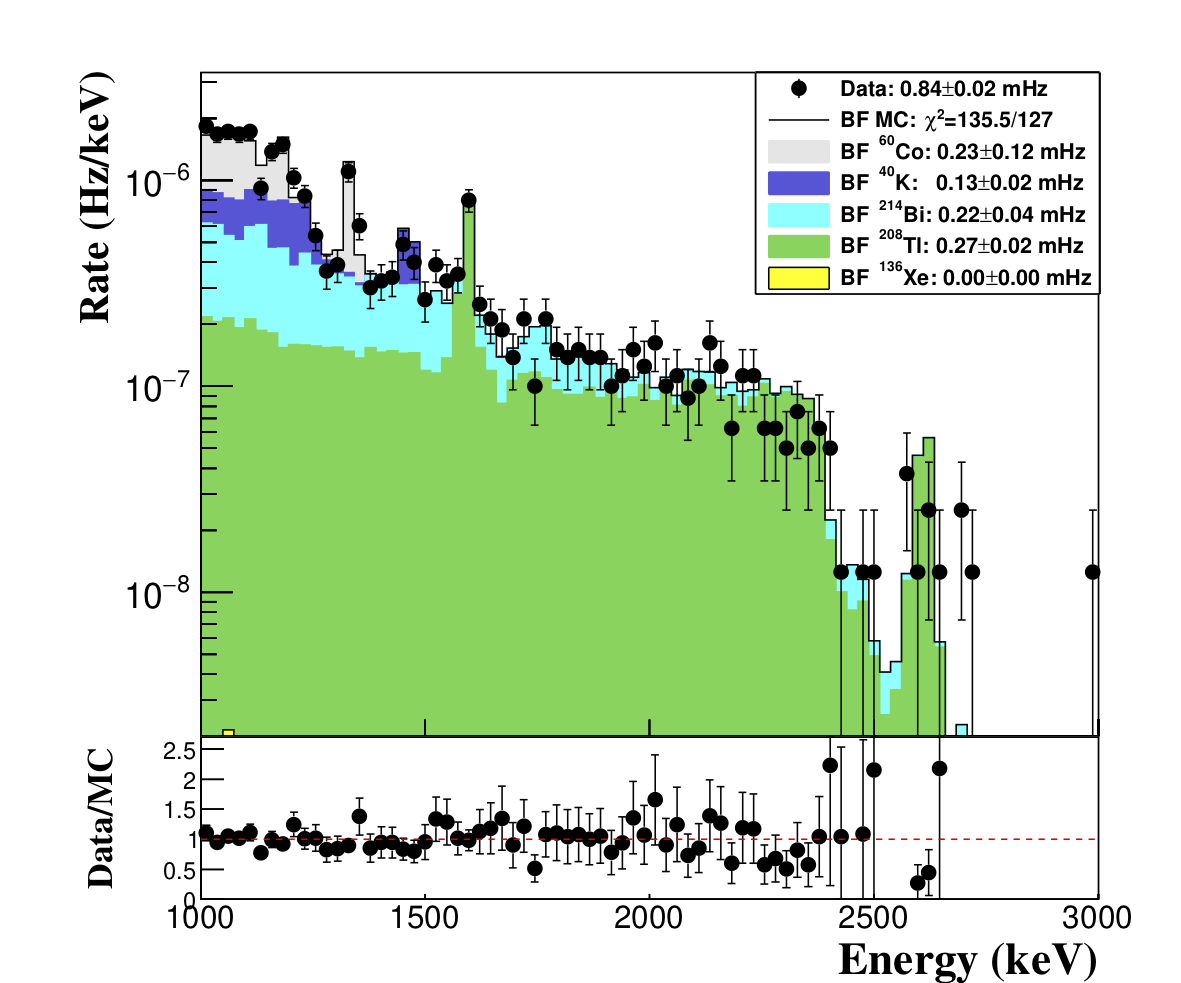}
\includegraphics[width=0.45\textwidth]{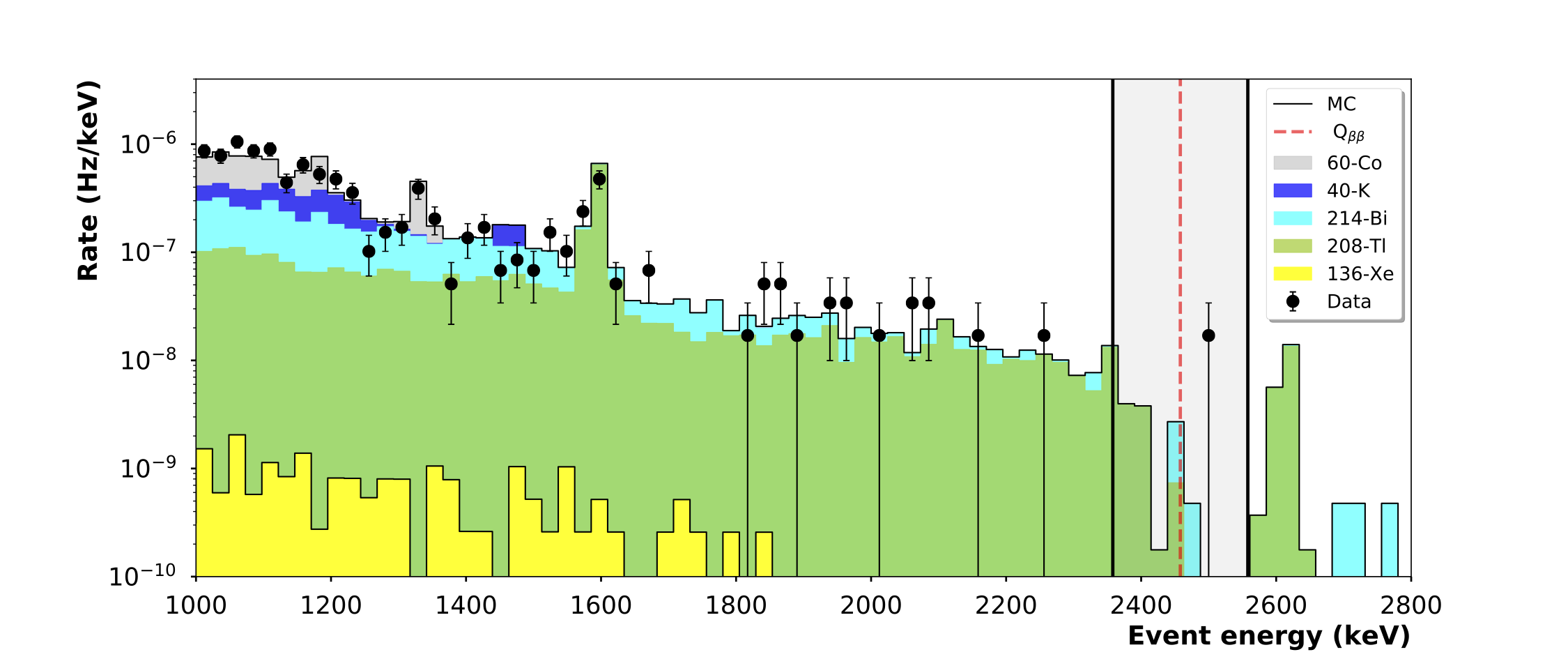}
\caption{Background spectrum measured by \NEW\ during Run IV (2018). The data fit well a model with four isotopes. Left panel: spectrum obtained only with fiducial cuts. Right panel: spectrum obtained adding topology cuts. Notice the depletion of the high energy region. }
\label{fig.newd}
\end{figure}

Operation of the detector during Run IV and V has shown excellent stability, with a very low spark rate and negligible leaks. {\em These two operational aspects were among the critical issues to be demonstrated by \NEW}. In addition, excellent electron lifetime has been achieved, in excess of \SI{5}{\milli\second} or 10 times the maximum drift length of the apparatus. Figures \ref{fig.csth},  \ref{fig.topo} and \ref{fig.newd} summarize some of the most relevant results:

$\bullet${\bf Energy resolution at \Qbb\ better than 1\% FWHM}

During Run IV and V enough statistics were acquired to measure the energy resolution at the energy of the \TL\ photopeak, which is very close to $\Qbb = \XenonQbb$. The result is shown in figure~\ref{fig.csth}. The fit to the \THO\ photopeak yields a resolution of 1\% FWHM in the full fiducial region~\citep{Renner:2019pfe}. {\em This result establishes \NEW\ as the xenon-based detector with the best energy resolution in the World}.

$\bullet${\bf Performance of the topological signature studied from the data themselves}

Using events near the \TL\ double escape peak at 1593~keV, it is possible to statistically separate a sample of signal-like, double-electron events (induced by pair production interactions) from a sample of background-like, single-electron events (induced by Compton interactions) (figure \ref{fig.topo}, left panel). The procedure can be carried out in data and in Monte Carlo, allowing a comparison between both of them. The right panel of figure~\ref{fig.topo} shows 
the signal efficiency versus the background acceptance for the two-electron identification algorithm, 
 The figure  shows good agreement between both datasets. A two-electron identification efficiency of 74\% for a background acceptance of 22\% (that is, 78\% background suppression) is measured at 1593~keV~ \citep{Ferrario:2019kwg,}, improving the results obtained in \citep{Ferrario:2015kta}. At \Qbb\ the discrimination improves, given the longer extensions of the tracks, which allows better separation of the blobs. A signal efficiency of 72\% and a background acceptance of 13\% is obtained. 

$\bullet${\bf Low background run}

Figure ~\ref{fig.newd}  shows a comparison between the predicted background rate as a function of energy and the one measured during Run IV, after a 34.5~d exposure and fiducial cuts. The expected nominal (pre-fit) background budget in \NEW\ is derived from a detailed background model accounting for different isotopes and detector volumes. The model relies on the extensive radiopurity measurements campaign conducted by the NEXT collaboration. The predicted (post-fit) background shown in the figure shows that the background in \NEW\ can be described in terms of just four isotopes.   Furthermore, the \NEW\ data acquired in Run IV permits a tuning of the background model from the data themselves~\citep{NewRadiogenic}.

The \bbonu\ backgrounds are also evaluated in an energy window around the Q$_{\bb}$ of \Xe{136} (2458 keV). A {\it loose \bbonu selection} is defined as the topological selection plus a Q$_{\bb}\pm$100~keV event energy requirement. Although this energy region is not representative of the $\sim$1\% FWHM energy resolution of the detector, it provides a statistically meaningful data/MC comparison using only 37.86~days of Run-IVc data and avoids the 2615~keV \Tl{208} photo-peak. This is the area shown by the light grey band in figure ~\ref{fig.newd}. One event is found to pass the loose \bbonu\ cuts in the entire Run-IVc period, in agreement with a MC expectation of (0.75$\pm$0.12$_{\rm stat}\pm$0.02$_{\rm syst}$) events. This provides a validation of the background model also in the \bbonu region of interest. 

\section{The \NEXT\ detector}
\label{sec.next100}

\begin{figure}[htb!]
\centering
\includegraphics[width=0.80\textwidth]{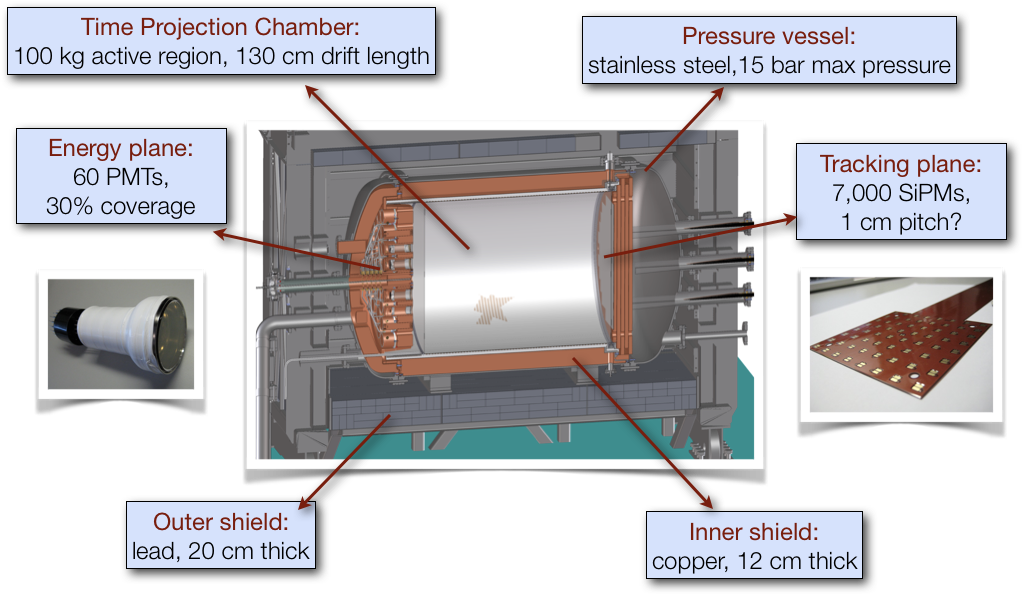}

\caption{\small The \NEXT\ detector.}
\label{fig.next-100}
\end{figure}

\begin{figure}[htb!]
\centering
\includegraphics[width=0.85\textwidth]{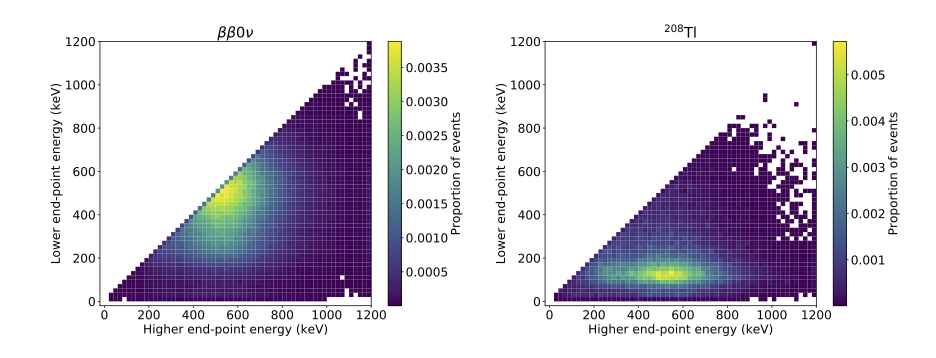}
\includegraphics[width=0.47\textwidth]{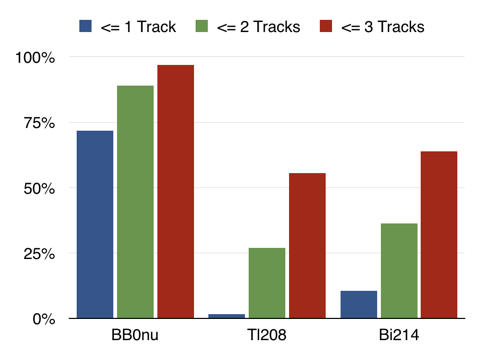} 
\includegraphics[width=0.40\textwidth]{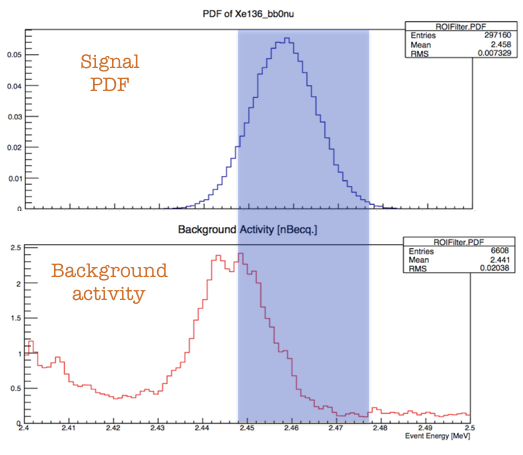}
\caption{\small Left: Top panel. The separation between blob energies for signal (\bbonu MC events) and \TL\ background (Monte Carlo events). Left bottom panel: track multiplicity for signal and background in NEXT-100. Right bottom panel: The ROI is optimized by maximizing a figure-of-merit that separates the signal from the closest background (\BI\ events). }
\label{fig.next-selec}
\end{figure}

The \Next\ detector is an asymmetric high-pressure xenon electroluminescent (\HPXeEL) TPC shown schematically in figure~\ref{fig.next-100}. The fiducial region is a cylinder of \NextTpcDiameter\ diameter and \NextTpcLength\ length, (\NextFiducialVolume\ fiducial volume) holding a mass of \NextFiducialMass\ of xenon gas enriched at \XeEnrichment\ in \XE, and operating at \NextPressure.  The energy plane (EP), featuring \NextNumberOfPMT\ PMTs.  The tracking plane (TP) features an array of \NextNumberOfSiPM\ SiPMs.  \Next\ scales up \NEW\ by roughly a factor 2.5 in longitudinal dimensions.

Figure~\ref{fig.next-selec} illustrates the handles that permit the separation of signal and background. These are: 

$\bullet${\bf Discrimination between 2 electrons and 1 electron}:
the top panel shows the energy distribution of the two extremes of the tracks (blobs). Signal events are characterized by two blobs of roughly the same energy, while for the background the energy of the less energetic blob is considerably smaller. This topological separation, demonstrated by \NEW\ can be further improved in the larger \Next\ detector. 

$\bullet${\bf Multiplicity}: the bottom left panel shows the track multiplicity for signal and background. Imposing a single track selects 75\% of the signal, while reducing the background by one order of magnitude. This is due to the larger bremsstrahlung of the energetic, single electrons that constitute the background, plus the photons associated to the production of the photopeaks. 

$\bullet${\bf Energy resolution}: the bottom right plane shows an example of ROI optimization, to maximize signal efficiency and reject the closest background, due to \TL. Energy resolution is essential here, given the close proximity of the \TL\  peak. 

\begin{figure}[htb!]
\centering
\includegraphics[width=0.51\textwidth]{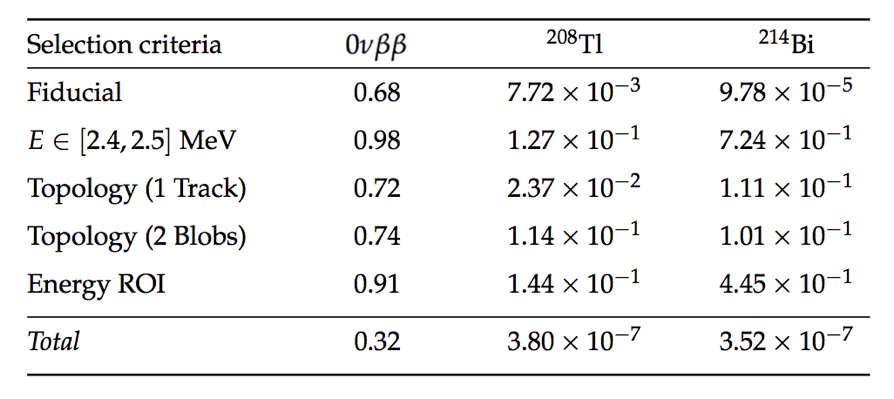}
\includegraphics[width=0.48\textwidth]{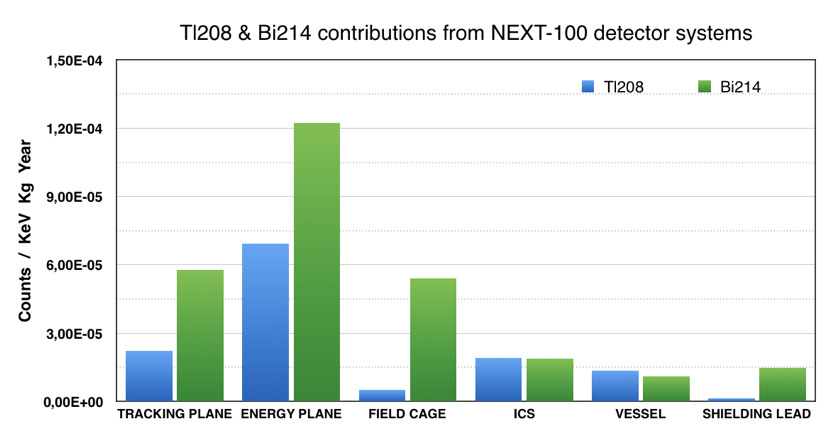}
\caption{\small Left panel: acceptance for signal events and rejection factors for the main radioactive background sources in NEXT-100. Right panel: background contributions in \ckky\ from the main subsystems of \Next.}
\label{fig.next-reje}
\end{figure}

 The left panel of figure~\ref{fig.next-reje} shows the acceptance for signal events and rejection factors for the main radioactive background sources in NEXT-100. The overall efficiency is 32\% for a rejection close to $3 \cdot 10^{-7}$~for the two main radioactive isotopes. The right panel shows the
 background contributions (in \ckky) from the main subsystems of \Next. As it can be seen the main sources are the energy plane (measured activity of the PMTs), the tracking plane (due to the activity of the circuit boards of the SiPMs) and the field cage, which however, uses a limit on the radioactivity of the HD polystyrene (no positive value is found from our current measurements). 
The over background rate is  \SI{4E-4}{\ckky} \citep{Martin-Albo:2015rhw}. This translates into a projected background in the ROI for \Next\ of $<$0.7~counts~yr\ensuremath{^{-1}}, with the leading background sources being the PMTs \citep{GomezCadenas:2017vp}. Operation of \Next\ is expected to start in 2020. This \Tonu\ sensitivity would translate into a \mbb\ reach as low as 60~meV, a value reachable if one assumes a favorable estimate for the nuclear matrix element.

In addition of its discovery potential, which is expected to be competitive with the best of the current generation of \bbonu\ experiments, NEXT-100, which is expected to be a virtually background-free experiment at the 100 kg scale, will be a demonstrator of the technology for the ton-scale.  

\section{Exploring the inverted hierarchy with NEXT}
\label{sec.ntk}

\subsection{NEXT-HD}
The \HPXeEL\ technology can be scaled up to multi-tonne target masses with some immediate gains due to improved material surface area to fiducial volume. Moreover, background will be reduced further by introducing several new technological advancements \citep{Cadenas:2017vcu}, for example: a)  the replacement of PMTs (which are the leading source of background in \Next) with SiPMs, which are intrinsically radiopure, resistant to pressure and able to provide better light collection ---notice that the radioactive budget of the tracking plane is due to the kapton circuit boards that will be replaced by ultra-low-background quartz circuits---; b) operation of the detector at cooler temperatures, in order to reduce the dark count rate of the SiPMs; and, c) optimization of the topological signature performance \citep{Renner:2017ey}; d) operation of the detector with low diffusion mixtures \citep{Henriques:2017rlj, Felkai:2017oeq, McDonald:2019fhy}, resulting in better position resolution and, thus, improving the performance of the topological signature. 
 We call this incremental approach to ton-scale \HPXeEL\ detectors "high definition" (HD).  The target of the HD technology is to reduce the specific background rate of \Next\ by at least one order of magnitude (thus counting less than background one event per ton per year of exposure). A NEXT-HD module with a mass in the ton range will be able to improve by more than one order of magnitude the current limits in \Tonu, thus exceeding $\Tonu > 10^{27}$~yr. 
 \subsection{NEXT-BOLD}
 
A more radical approach could implement a barium tagging sensor capable of detecting with high efficiency the presence of the \Bapp\ ion produced in the \XE\ \bbonu\ decay.

\begin{equation}
\XE \rightarrow 2\, e^- + \Bapp \, (+2\nu).
\label{eq:bbonu}
\end{equation}

We call this disruptive approach``Barium iOn Light Detection'' (BOLD).

Natural radioactivity does not create \Bapp\ ions. If the double positive barium ion could be detected {\it in coincidence}\footnote{It would be, in fact, delayed coincidence, since the ion will take about ten times longer to drift to the cathode than the ionization electrons to drift to the anode.} with the two-electron signal (and for events in the narrow ROI allowed by the good energy resolution), the expected background rate drops to zero. 

Is it possible to detect (``tag'') the \Bapp\ ion produced in the xenon \bbonu\ decay? The first question to discern is whether the single \Bapp\ ion will survive the trip to the cathode. The answer appears to be a cautious yet \citep{Bainglass:2018odn}. Assuming that this is the case, one needs to instrument the cathode in order to implement a system to focus the ion into a small sensor region. The ion could be focused towards the sensors using radio frequency \citep{ARAI201456}, a method already demonstrated at large scales \citep{Gehring2016221} and for barium transport in HPXe \citep{Brunner:2014sfa}. 

All the above, while technically challenging, appears reasonably within the range of existing technology. {\it The challenge of \Bapp\ tagging and the key to a background free experiment, is, therefore,  to build sensors capable of detecting a single ion with high efficiency}.

\subsection{SMFI for Barium tagging:}
 
 The possibility of using SMFI as the basis of molecular sensors for Barium tagging was proposed in \citep{nygrenbata, Jones:2016qiq}, followed,  shortly after, by a proof of concept which managed to resolve individual 
\Bapp\ ions on a scanning surface using an SMFI-based sensor \citep{McDonald:2017izm}. The SMFI sensor concept used a thin quartz plate with surface-bound fluorescent indicators, continuously illuminated with excitation light and monitored by an EM-CCD camera. 
The chosen indicator was FLUO-3, a common fluorophore,  suspended in polyvinyl alcohol (PVA). This partially emulated the conditions in a HPXe TPC detector, where the ions will drift to the sensor plate and adhere to fluorophores immobilized there. 

And yet, the setup of \citep{McDonald:2017izm} would not work in a \HPXeEL. The gas in such a detector must be  free of impurities ---including water and CH compounds, \eg alcohol--- at the part-per-trillion level. This alone would rule out a PVA-based sensor. Furthermore, the target will need to be densely populated by sensor molecules, to ensure maximum ion-capture efficiency, while the target of \citep{McDonald:2017izm} was sparsely populated in order to facilitate the observation of individual chelated molecules. 

A realistic sensor for barium tagging in a \HPXeEL\, thus, needs to be based on a monolayer of indicators which able to fluoresce in dry medium. As it turns out, the fluorescence of FLUO-3 and related molecules is heavily suppressed in dry medium \citep{Byrnes:2019jxr}, making them unsuitable for this application. In addition of shining in dry gas, an indicator for barium tagging has to yield a high signal to noise ration between chelated and unchelated molecules. In another words, the emission spectrum of chelated molecules must be overlap as little as possible with that of unchelated molecules. Last but not least, it is desirable, although not compulsory, that the emission spectrum is in the visible, to increase detection efficiency. 

The NEXT collaboration has already succeeded in synthesizing new molecules able to fluoresce in dry medium
\citep{Thapa:2019zjk, FIB}. The R\&D for NEXT-BOLD includes the characterization of those molecules (and the potential synthesis of new indicators), the development of molecular mono-layers, and the development of suitable laser optics. A second proof-of-concept in dry medium is under way. 

 \subsection{Sensitivity of NEXT.}
\begin{figure}[htb!]
\centering
\includegraphics[width=0.45\textwidth]{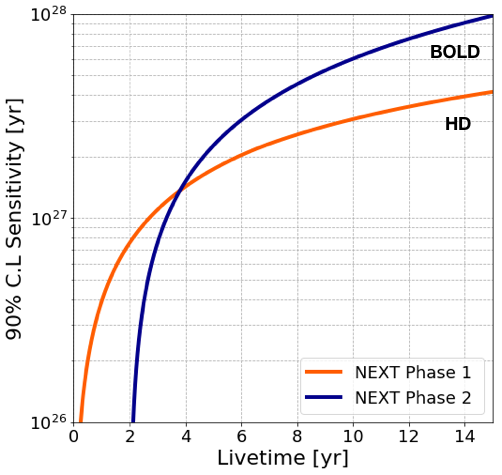}

\caption{\small Sensitivity of NEXT.}
\label{fig.next-sensi}
\end{figure}

The next-generation \HPXeEL\ could deploy one or more modules, with a mass approaching one ton per module and based in either HD or BOLD approaches, depending on readiness of the technologies. A possible scenario would be to deploy a first HD module followed by a second, BOLD module.  The target background of the \Ntk-HD module is \SI{< 1}{\ev~tonne\ensuremath{^{-1}}~yr\ensuremath{^{-1}}}, allowing the experiment to reach a half-life sensitivity of \SI{2E27}{\yr} with an exposure of \SI{6}{\tonne\cdot\yr} (figure ~\ref{fig.next-sensi}). With \Bapp-tagging\ a truly background-free experiment capable of exploring the physical parameter space even further (or faster) would be possible. A sensitivity of \Tonu=\SI{8E27}{\yr} after a \SI{10}{\tonne\cdot\yr} exposure would be expected in  this case.

\section{Summary and outlook}
\label{sec.conclu}
After ten years of development, the \HPXeEL\ technology for \bbonu\ searches is becoming mature. The \NEW\ detector, a radiopure apparatus deploying 5 kg of enriched xenon is currently taking data at the LSC, in Spain, and has demonstrated all the main aspects of the technology. NEXT-100 will start operations next year and is expected to be a background-free experiment at the 100 kg scale. Two complementary approaches, NEXT-HD and NEXT-BOLD, are being considered for the next generation experiments needed to explore the inverse hierarchy. If the BOLD approach can be realized, then NEXT would become a truly background free experiment, and thus the technology could be extrapolated optimally to very large masses. 

\section*{Acknowledgments}

The author would like to warmly thank the NEXT collaboration whose collective effort is represented in this status report. This research has been supported by the European Research Council (ERC) under the Advanced Grant 339787-NEXT; the Ministerio de Econom\'ia y Competitividad of Spain under grants FIS2014-53371-C04 and the Severo Ochoa Program SEV-2014-0398; the GVA of Spain under grant PROMETEO/2016/120;


\bibliographystyle{JHEP}
\bibliography{moriond2019}

\end{document}